\documentstyle[12pt]{article}
\def\g {\gamma}

\def\J {$J/\psi\;\;$}

\def\Y {$\Upsilon\;\;$}
\def\F {{\cal F}}
\def\R {{\cal R}}

\def\cc {$c\bar{c}\;\;$}
\def\qq {$Q\bar{Q}\;\;$}
\def\k {$k_T$-factorization~}
\renewcommand{\thefootnote}{\alph{footnote}}
\def\ap#1#2#3   {#3 {\sl Ann. Phys. (NY)}         {\bf#1}   #2}
\def\apj#1#2#3  {#3 {\sl Astrophys. J.}           {\bf#1}   #2}
\def\apjl#1#2#3 {#3 {\sl Astrophys. J. Lett.}     {\bf#1}   #2}
\def\app#1#2#3  {#3 {\sl Acta. Phys. Pol.}        {\bf#1}   #2}
\def\cpc#1#2#3  {#3 {\sl Computer Phys. Comm.}    {\bf#1}   #2}
\def\dum#1#2#3  {#3 {~}                           {\bf#1}   #2}
\def\epjc#1#2#3 {#3 {\sl Eur. Phys. J.} C         {\bf#1}   #2}
\def\err#1#2#3  {#3 {\it Erratum}                 {\bf#1}   #2}
\def\etal       {{\it et al.~}}
\def\ib#1#2#3   {#3 {\it ibid.}                   {\bf#1}   #2}
\def\jcp#1#2#3  {#3 {\sl J. Comp. Phys.}          {\bf#1}   #2}
\def\jmp#1#2#3  {#3 {\sl J. Math. Phys.}          {\bf#1}   #2}
\def\ijmp#1#2#3 {#3 {\sl Int. J. Mod. Phys.}      {\bf#1}   #2}
\def\jpg#1#2#3  {#3 {\sl J. Phys. } G             {\bf#1}   #2}
\def\mpl#1#2#3  {#3 {\sl Mod. Phys. Lett.}        {\bf#1}   #2}
\def\nat#1#2#3  {#3 {\sl Nature (London)}         {\bf#1}   #2}
\def\ncim#1#2#3 {#3 {\sl Nuovo Cimento}           {\bf#1}   #2}
\def\nim#1#2#3  {#3 {\sl Nucl. Instr. Meth.}      {\bf#1}   #2}
\def\np#1#2#3   {#3 {\sl Nucl. Phys.}             {\bf#1}   #2}
\def\npa#1#2#3  {#3 {\sl Nucl. Phys.} A           {\bf#1}   #2}
\def\npb#1#2#3  {#3 {\sl Nucl. Phys.} B           {\bf#1}   #2}
\def\pan#1#2#3  {#3 {\sl Phys. At. Nuclei}        {\bf#1}   #2}
\def\pl#1#2#3   {#3 {\sl Phys. Lett.}             {\bf#1}   #2}
\def\plb#1#2#3  {#3 {\sl Phys. Lett.} B           {\bf#1}   #2}
\def\prep#1#2#3 {#3 {\sl Phys. Rep.}              {\bf#1}   #2}
\def\prev#1#2#3 {#3 {\sl Phys. Rev.}              {\bf#1}   #2}
\def\prc#1#2#3  {#3 {\sl Phys. Rev.} C            {\bf#1}   #2}
\def\prd#1#2#3  {#3 {\sl Phys. Rev.} D            {\bf#1}   #2}
\def\prev#1#2#3 {#3 {\sl Phys. Rev.}              {\bf#1}   #2}
\def\prl#1#2#3  {#3 {\sl Phys. Rev. Lett.}        {\bf#1}   #2}
\def\prs#1#2#3  {#3 {\sl Proc. Roy. Soc.}         {\bf#1}   #2}
\def\ptp#1#2#3  {#3 {\sl Prog. Theor. Phys.}      {\bf#1}   #2}
\def\ps#1#2#3   {#3 {\sl Physica Scripta}         {\bf#1}   #2}
\def\slp#1#2#3  {#3 {\sl Rev. Mod. Phys.}         {\bf#1}   #2}
\def\rpp#1#2#3  {#3 {\sl Rep. Prog. Phys.}        {\bf#1}   #2}
\def\sjnp#1#2#3 {#3 {\sl Sov. J. Nucl. Phys.}     {\bf#1}   #2}
\def\spj#1#2#3  {#3 {\sl Sov. Phys. JETP}         {\bf#1}   #2}
\def\spjl#1#2#3 {#3 {\sl Sov. JETP Lett.}         {\bf#1}   #2}
\def\spu#1#2#3  {#3 {\sl Sov. Phys.-Usp.}         {\bf#1}   #2}
\def\zp#1#2#3   {#3 {\sl Zeit. Phys.}             {\bf#1}   #2}
\def\zpc#1#2#3  {#3 {\sl Zeit. Phys.} C           {\bf#1}   #2}
\begin{document} 
\begin{large} \begin{bf} \noindent
      Deep inelastic \J production at HERA\\
      in the \k approach \\
      and its consequences on the nonrelativistic QCD
\end{bf} \end{large}
~\\~\\
 { S.~P.~Baranov}$~^{1,}$%
\footnote{\parbox[t]{10cm}{Electronic address: baranov@sci.lebedev.ru}},
 { N.~P.~Zotov}$~^{2,}$%
\footnote{\parbox[t]{10cm}{Electronic address: zotov@theory.sinp.msu.ru}}
~\\~\\
\begin{small}
\begin{tabular}{rp{15cm}}
$~^1$&{\rm
P.N.~Lebedev~Physics~Institute,~Leninsky prosp.~53,~Moscow~119991,~Russia}\\
$~^2$&{\rm D.V.~Skobeltzyn Institute of Nuclear Pysics,}\\
    ~&{\rm M.V.~Lomonosov Moscow State University,~Moscow~119299,~Russia}
\end{tabular}\\~\\
Short title: Deep inelastic \J production at HERA\\
PACS numbers(s):  12.38.Bx, 13.85.Ni, 14.40.Gx \\
\vspace*{5mm}~\\
{\bf Abstract.}
In the framework of the $k_T$-factorization approach, we analyse the inclusive
and inelastic production of \J particles in deep inelastic $ep$ scattering.
We take into account both colour-singlet and colour-octet production channels.
We inspect the sensitivity of theoretical predictions to the choice of model
parameters. Our theoretical results agree reasonably well with recent
experimental data collected by the collaboration H1 at HERA.
\\~\\
\end{small}
\renewcommand{\thefootnote}{\arabic{footnote}} \setcounter{footnote}{0}

\noindent {\bf  1~Introduction}\\

Over the last decade, investigation of the \J production mechanisms in
hadron-hadron and lepton-hadron collisions continues to attract significant
attention from both theoretical and experimental sides.
The puzzling history of \J traces back to the early 1990s, when the
measurements \cite{CDF0}-\cite{CDF3} of the \J and \Y hadroproduction
cross sections revealed a more than one order-of-magnitude discrepancy with
theoretical expectations \cite{Chang}-\cite{Krase}. This fact has induced
extensive theoretical activity. In particular, it led to the introduction
of a new production mechanism, the so called colour-octet model \cite{BraFle}-%
\cite{Sridha}. Since then, the colour-octet model has been believed to give
the most likely explanation of the quarkonium production phenomena, although
there are also some indications that it is not working well \cite{Mizuk}.
One of the problems is connected with the photoproduction data \cite{H1},
\cite{ZEUS} where the contribution from the colour-octet mechanism is
unnecessary or even unwanted \cite{CacKra}-\cite{Beneke} as the experimental
results can be described within the colour-singlet model alone (if the
next-to-leading-order contributions are taken into account) \cite{Kraem}.
Another difficulty refers to the \J spin alignment. If, as expected, the
dominant contribution comes from the gluon fragmentation into an octet \cc
pair, the mesons must have strong transverse polarization \cite{Beneke}-%
\cite{Lee}. This is in disagreement with the data \cite{E537}, \cite{CDF},
\cite{Glad}, \cite{newH1} which point to unpolarized or even longitudinally
polarized mesons.

A different strategy is represented by the \k approach
\cite{Gribov}-\cite{Collins}
(see also recent review \cite{Bo} and references therein). In this approach,
one focuses on the resummation of ``small $x$'' logarithms (i.e., the terms
$[\mbox{ln}(\mu^2/\Lambda^2)\,\alpha_s]^n$,
$[\mbox{ln}(\mu^2/\Lambda^2)\,\mbox{ln}(1/x)\,\alpha_s]^n$, and
$[\mbox{ln}(1/x)\,\alpha_s]^n$) to all orders in $n$ and takes into account
the effects of finite transverse momenta of partons. The resummation results
in the ``unintegrated'' parton distributions which generalize the QCD
factorization beyond the collinear approximation.


The effects of the initial gluon transverse momentum and gluon off-shellness
were shown to have an important impact on the \J production properties in the
photon-gluon fusion colour-singlet model \cite{jbfkl}, \cite{Artem}.
Several attempts have been made in the literature to incorporate the
colour-octet model within the \k scheme \cite{Teryaev}, \cite{Chao2}.
An extensive analysis of the production of $J/\psi$, $\chi_c$, and $\Upsilon$
mesons in $p\bar{p}$ collisions at the Fermilab Tevatron has been recently
presented in Ref. \cite{j_sha}. Our paper is a logical continuation of this
line. Here, we concentrate on the inclusive and inelastic production of \J
particles by virtual photons in deep inelastic $ep$ scattering at HERA. The
calculations are based on the \k approach and on the nonrelativistic QCD
formalism where we take into account both colour-singlet and colour-octet
channels.

The outline of the article is as follows. First, in Sec. 2, we briefly recall
the basic principles of the colour-singlet and colour-octet models and explain
their extension to the $k_T$-factorization approach. The necessary technical
details are described in Sec. 3. The numerical results followed by a
discussion are displayed in Sec. 4. Finally, the concluding remarks are
collected in Sec. 5.\\

\noindent {\bf  2~Theoretical framework}\\

In the framework of the colour-singlet approach \cite{Chang}-\cite{Krase},
the production of any heavy meson is described as the perturbative production
of a colour-singlet \qq pair in a state with properly arranged the quantum
numbers, according to the quarkonium state under consideration. For the
production of \J particles, the relevant partonic subprocess is
\begin{eqnarray}
\g+g &\to& ~^3S_1[1]+g, \label{psig}
\end{eqnarray}
where we follow the standard spectroscopic notations, and the number in the
brackets stands for the colour representation of the \cc pair.
The corresponding Feynman diagrams are depicted in Fig. 1(a)
(assuming also five possible permutations of the photon and gluon lines).
The formation of a meson from the quark pair is a nonperturbative process.
Within the nonrelativistic approximation which we are using, this probability
reduces to a single parameter related to the meson wave function at the origin
$|\R(0)|^2$, which is known for the \J and \Y families from the measured
leptonic decay widths.

In addition to the above, we consider the colour-octet production scheme
\cite{BraFle}-\cite{ChoLei}, which implies that the heavy \qq pair is
perturbatively created in a hard subprocess as an octet colour state and
subsequently evolves into a physical quarkonium state via emitting soft
(nonperturbative) gluons, which may be interpreted as a series of
``classical'' colour-dipole transitions.
Although these transition probabilities can, in principle, be expressed
in terms of field operators and therefore calculated, no such calculation
exists to date. Thus, the transition probabilities remain free parameters,
which are assumed to obey a definite hierarchy in powers of $v$, the relative
velocity of the quarks in the bound system under study. This freedom is
commonly used to estimate the colour-octet parameters by adjusting them to
experimental data.

In the case when the colour-octet \qq state is allowed, there appear 
additional contributions from the diagram of Fig. 1(a) and another set of
diagrams shown in Figs. 1(b) and 1(c) (including all possible permutations). 
The graphs shown in Figs. 1(a) and 1(b) correspond to the following partonic 
subprocesses
\begin{eqnarray}
\g+g &\to& ~^1S_0[8]+g, \label{1S0g}\\
\g+g &\to& ~^3S_1[8]+g, \label{3S1g}\\
\g+g &\to& ~^3P_J[8]+g, \label{3PJg}
\end{eqnarray}
while the ones shown in Fig. 1(c) refer to the subprocesses
\begin{eqnarray}
\g+g &\to& ~^1S_0[8], \label{1S0}\\
\g+g &\to& ~^3P_J[8]. \label{3PJ}
\end{eqnarray}
(The production of $^3S_1[8]$ state in a $2\to 1$ photon-gluon fusion process
is forbidden by the colour and charge parity conservation.)
Although the $2\to 2$ subprocesses (\ref{1S0g})-(\ref{3PJg}) are of formally
subleading order in $\alpha_s$ in comparison with (\ref{1S0})-(\ref{3PJ}),
their role cannot be regarded as small. Since they contribute to very different
regions of the phase space, they even can dominate over the $2\to 1$
subprocesses under the experimental conditions. We will discuss this in more
detail in Sec. 4. Note also that the $2\to 2$ subprocesses are indispensable
for the inelastic events (i.e., the events with large final state hadron mass).

 The colour-octet matrix elements (usually denoted in the literature as     %
 $\langle 0|{\cal O}_8|0\rangle$) responsible for the nonperturbative       %
 transitions in (\ref{1S0g})-(\ref{3PJ}) are related to the fictituous      %
 colour-octet wave functions, that are used in calculations in place of     %
 the ordinary colour-singlet wave functions:                                %
 $$ 
 \langle 0|{\cal O}_8|0\rangle =                                            %
 \frac{9}{2\pi}|\R_8(0)|^2=\frac{9}{2\pi}\,4\pi\,|\Psi_8(0)|^2.             %
 $$ 
 This equation applies to all $S$-wave colour-octet states, and             %
 a similar relation holds for $\R'_8(0)$ and $\Psi'_8(0)$ if the $P$-wave   %
 colour-octet states are involved.                                          %
 For the sake of uniformity, we will be consistently using the notation     %
 in terms of $\R(0)$ and $\R'(0)$ for both colour-singlet and colour-octet  %
 contributions.                                                             %

A generalization of the above formalism to the $k_T$-factorization approach
implies two important steps. These are the introduction of noncollinear gluon
distributions (which we show here) and the modification of the gluon spin
density matrix in the parton level matrix elements (which we explain in the
next section).

In the numerical analysis, we have tried two different sets of $k_T$-dependent
gluon densities. In the approach of Ref. \cite{Bluem} based on a leading-order
perturbative solution of the Balitsky-Fadin-Kuraev-Lipatov (BFKL) \cite{BFKL}
equation, the unintegrated gluon density $\F_g(x,k_{T}^2,\mu^2)$ is calculated
as a convolution of the ordinary (collinear) gluon density $G(x,\mu^2)$ with
universal weight factors:
\begin{equation} \label{conv}
 {\cal F}_g(x,k_{T}^2,\mu^2) = \int_x^1
 {\cal G}(\eta,k_{T}^2,\mu^2)\,
 \frac{x}{\eta}\,G\left(\frac{x}{\eta},\mu^2\right)\,d\eta,
\end{equation}
\begin{equation} \label{J0}
 {\cal G}(\eta,k_{T}^2,\mu^2)=\frac{\bar{\alpha}_s}{\eta\,k_{T}^2}\,
 J_0(2\sqrt{\bar{\alpha}_s\ln(1/\eta)\ln(\mu^2/k_{T}^2)}),
 \qquad k_{T}^2<\mu^2,
\end{equation}
\begin{equation}\label{I0}
 {\cal G}(\eta,k_{T}^2,\mu^2)=\frac{\bar{\alpha}_s}{\eta\,k_{T}^2}\,
 I_0(2\sqrt{\bar{\alpha}_s\ln(1/\eta)\ln(k_{T}^2/\mu^2)}),
 \qquad k_{T}^2>\mu^2,
\end{equation}
where $\mu$ is the factorization scale,
$J_0$ and $I_0$ stand for Bessel functions (of real and imaginary
arguments, respectively), and $\bar{\alpha}_s=3\alpha_s/\pi$. In the leading-%
order approximation, the parameter $\bar{\alpha}_s$ is connected to the Pomeron
intercept $\alpha(0)=1+\Delta$, with $\Delta=\bar{\alpha}_s\,4\ln{2}$. We use
the value of $\Delta=0.35$ as it is accepted in many other our papers
(\cite{BJJPZ} and references therein).

Another parametrization is based on the
Dokshitzer-Gribov-Lipatov-Altarelli-Parisi (DGLAP) \cite{DGLAP} evolution
equation. This approach was originally proposed in \cite{Gribov}, \cite{BFKL}
and is now frequently discussed in the literature \cite{Shab}, \cite{Nikol}.
It recalls the kinematic
relation between the virtuality $q^2$ and the transverse momentum $k_T$ of a
parton: $q^2 = k_T^2/(1-x)$. Consequently, the ordinary gluon density
$G(x,q^2)$ may be considered as giving the $k_T^2$ distribution also.
In this approach, the unintegrated gluon density is derived from the
``collinear'' density by simply differentiating it with respect to $q^2$:
\begin{equation} \label{dglap}
{\cal F}_g(x,k_T^2,\mu^2=k_T^2) =\frac{d}{dq^2}\,G(x,q^2)|_{q^2=k_T^2}.
\end{equation}
As the BFKL and DGLAP equations are known to collect different logarithms,
we find it worth exploring the numerical consequences of this difference.
For consistency, the same leading-order (LO) GRV set \cite{GRV95} was used
in both cases as the input collinear density.\\

\noindent {\bf  3~Matrix elements and differential cross section}\\

\noindent
At first, we consider the relevant $2\to 2$ partonic subprocesses as they are
given by the photon-gluon fusion mechanism.
Let $k_1$, $k_2$, $k_3$, $p_{c}$, and $p_{\bar{c}}$ be the momenta of the
incoming virtual photon and gluon, the outgoing gluon and the outgoing heavy
(charmed) quark and antiquark, respectively; $\epsilon_1$, $\epsilon_2$, and
$\epsilon_3$ the photon and gluon polarization vectors; $m_c$ the quark mass,
$m_\psi=2m_c$, $k{=}k_2{-}k_3$, and $a$, $b$, and $c$ the eight-fold colour
indices of the incoming gluon, the outgoing gluon, and the (coloured) \cc state.
We also introduce the projection operator $J(S,L)$, which guarantees the
proper spin and orbital angular momentum of the \cc state under consideration.
Then, the photon-gluon fusion matrix elements read
\begin{eqnarray}
&&\hspace*{-1cm}{\cal M}_a(\g g\to\psi g)=
 tr \{\not\epsilon_1\,(\not p_c - \not k_1 + m_c)
      \not\epsilon_2\,(-\not p_{\bar c} - \not k_3 + m_c)
      \not\epsilon_3\,J(S,L)\} \nonumber \\ &&
    \times C_{\psi}\;tr\{T^aT^bT^c\}\;
    [k_1^2-2(p_c k_1)]^{-1} [k_3^2+2(p_{\bar c}k_3)]^{-1} \label{Ma}
    \mbox{\quad + 5 permutations},\\
&&\hspace*{-1cm}{\cal M}_b(\g g\to\psi g)=
 tr \{\not\epsilon_1\,(\not p_c - \not k_1 + m_c)
      \gamma_\mu\,J(S,L)\}\,i
    G^{(3)}(k_2,\epsilon_2,-k_3,\epsilon_3,-k,\mu) \nonumber \\ &&
    \times C_{\psi}\;f^{abd}\,tr\{T^cT^d\}\;
    [k_1^2-2(p_c k_1)]^{-1} [k^2]^{-1} \label{Mb}
    \mbox{\quad + 1 permutation}.
\end{eqnarray}
In the above expression, $G^{(3)}$ is related to the standard QCD
three-gluon coupling
\begin{equation}
    G^{(3)}(p,\lambda,~q,\mu,~k,\nu)=
               (q-p)^{\nu}g^{\lambda \mu}
              +(k-q)^{\lambda}g^{\mu \nu}
              +(p-k)^{\mu}g^{\nu \lambda}.
\end{equation}
The factor represented by the $SU(3)$ generator matrix $T^c$ has to be
replaced by the unity matrix if the outgoing \cc state is a colour
singlet. The coefficient $C_{\psi}$ stands for the normalization of the
\cc colour wave function and is equal to $1/\sqrt{3}$ and $1/2$ for the
singlet and octet states, respectively.

The projection operator $J(S,L)$ reads for the different spin and orbital
angular momentum states \cite{Chang}-\cite{Krase}:
\begin{eqnarray}
\label{j00} J(^1S_0)\equiv
J(S{=}0,L{=}0)&=&\gamma_5\,(\not{p_c}+m_c)/m_{\psi}^{1/2}, \\
\label{j10} J(^3S_1)\equiv
J(S{=}1,L{=}0)&=&\not{\epsilon(S_z)}\,(\not{p_c}+m_c)/m_{\psi}^{1/2}, \\
\label{j11} J(^3P_J)\equiv
J(S{=}1,L{=}1)&=&
(\not{p_{\bar{c}}}-m_c)\,\not{\epsilon(S_z)}\,(\not{p_c}+m_c)/m_{\psi}^{3/2}.
\end{eqnarray}
States with various projections of the spin momentum onto the $z$ axis are
represented by the polarization vector $\epsilon(S_z)$.

In the nonrelativistic approximation which we are using, the relative
momentum $q$ of the quarks in the bound state is treated as a small
quantity. So, it is useful to represent the quark momenta as follows:
\begin{equation}
p_{c}=\frac{1}{2}p_{\psi}+q, \quad p_{\bar{c}}=\frac{1}{2}p_{\psi}-q,
\end{equation}
where $p_{\psi}$ is the momentum of the final state quarkonium.
The probability for the two quarks to form a meson depends on the bound
state wave function $\Psi(q)$. Therefore, we multiply the matrix elements
(\ref{Ma})--(\ref{Mb}) by $\Psi(q)$ and perform integration with respect to
$q$. The integration is performed after expanding the integrand around $q=0$:
\begin{equation}
{\cal M}(q)={\cal M}|_{q=0}+
q^\alpha\,(\partial{\cal M}/\partial q^\alpha)|_{q=0}+\,\,\dots\label{exp}
\end{equation}
Since the expressions for
${\cal M}|_{q=0}$, $(\partial{\cal M}/\partial q^\alpha)|_{q=0}$, etc., are
no longer dependent on $q$, they may be factored outside the integral sign.
A term-by-term integration of this series then yields \cite{Krase}:
\begin{eqnarray}
 &&\int \frac{d^3q}{(2\pi)^3}\,\Psi(q)=\frac{1}{\sqrt{4\pi}}\,\R(x=0),
 \label{r0} \\
 &&\int \frac{d^3q}{(2\pi)^3}\,q^\alpha\Psi(q)=
 -i\epsilon^\alpha(L_z)\,\frac{\sqrt{3}}{\sqrt{4\pi}}\,\R'(x=0),
 \label{r1}
\end{eqnarray}
etc., where $\R(x)$ is the spatial component of the wave function in the
coordinate representation (the Fourier transform of $\Psi(q)$). The first
term contributes only to $S$-waves, but vanishes for $P$-waves because
$\R_P(0)=0$. On the contrary, the second term contributes only to $P$-waves,
but vanishes for $S$-waves because $\R'_S(0)=0$.
States with various projections of the orbital angular momentum onto the
$z$ axis are represented by the polarization vector $\epsilon(L_z)$.

The polarization vectors $\epsilon(S_z)$ and $\epsilon(L_z)$ are defined
as explicit four-vectors. In the frame where the $z$ axis is oriented along
the quarkonium momentum vector, $p_{\psi}=(0,\,0,\,|p_{\psi}|,\,E_{\psi})$,
these polarization vectors read:
\begin{equation}
\epsilon(\pm 1)  = (1,\,\pm i,\,0,\,0)/\sqrt{2},\qquad
\epsilon(0) = (0,\,0,\,E_{\psi},\,|p_{\psi}|)/m_{\psi}.
\end{equation}
States with definite $S_z$ and $L_z$ can be translated into states with
definite total momentum $J$ and its projection $J_z$ using
the Clebsch-Gordan coefficients:
\begin{equation}
\epsilon^{\mu\nu}(J,J_z)=\sum_{L_z,S_z} \label{jz}
\langle 1,L_z;1,S_z|J,J_z\rangle\epsilon^{\mu}(L_z)\epsilon^{\nu}(S_z).
\end{equation}

As far as the gluon spin density matrix is concerned, one should take into
account that gluons generated in the parton evolution cascade do carry
non-negligeable transverse momentum and are off mass shell. One can trace a
clear analogy between the \k approach and the equivalent photon approximation
in QED showing that the polarization properties of the off-shell incoming
gluon are similar to those of the incoming virtual photon.
The off-shell photon spin density matrix is given by the full lepton tensor
\begin{equation} \label{Lmunu}
\overline{\epsilon^{\mu}\epsilon^{*\nu}}\sim
         L^{\mu\nu} = 8\,p^\mu p^\nu - 4(pk)\,g^{\mu\nu},
\end{equation}
where $p$ is the momentum of the beam particle, and $k$ is the momentum of the
emitted photon. A similar anzatz (Eq. (\ref{epsk}), see below) is used in the
$k_T$-factorization approach. Neglecting the second term in the right hand
side of (\ref{Lmunu}) in the small $x$ limit, $p\gg k$, one arrives at the
spin structure  $\overline{\epsilon^{\mu}\epsilon^{*\nu}}\sim p^\mu p^\nu$.
The latter can be rewritten in the form
\begin{equation} \label{epsk}
\overline{\epsilon^{\mu}\epsilon^{*\nu}}=k_T^{\mu}k_T^{\nu}/|k_T|^2,
\end{equation}
where we have represented the 4-momentum $k$ as $k=xp+k_T$ and applied
a gauge shift $\epsilon^\mu\to\epsilon^\mu-k^\mu/x$. This formula converges
to the usual $\sum\epsilon^{\mu}\epsilon^{*\nu}=-g^{\mu\nu}$ when $k_T\to 0$.
In the present calculations, we use Equ. (\ref{Lmunu}) for virtual photons,
and Equ. (\ref{epsk}) for off-shell gluons.
The expressions (\ref{Lmunu}) and (\ref{epsk}) merge with each other in the
ultrahigh energy limit. The effect of the second term in (\ref{Lmunu})
at HERA energies is found to be about 5 to 10 percent \cite{BJJPZ}.
As we will see in Section 4, the presence of longitudinal components in the
off-shell gluon spin density matrix has important impact on the quarkonium
polarization.
The final state gluon in (\ref{Ma})-(\ref{Mb}) is assumed on-shell,
$\sum{\epsilon_3^{\mu}\epsilon_3^{*\nu}}=-g^{\mu\nu}$.
The evaluation of the traces in Eqs. (\ref{Ma})-(\ref{Mb}) is straightforward
and is done using the algebraic manipulation system FORM \cite{FORM}.

To calculate the cross section of a physical process we have to multiply the
matrix elements squared by the gluon distribution functions and perform
integration over the final state phase space. The multiparticle phase space
$\prod d^3p_i/(2E_i)\,\delta^4(\sum p_{in}-\sum p_{out})$ is parametrized
in terms of transverse momenta, rapidities, and azimuthal angles:
$d^3p_i/(2E_i)$=$(\pi/2)dp_{iT}^2dy_i\,d\phi_i/(2\pi)$.
Let $\phi_1$ and $\phi_2$ be the azimuthal angles of the scattered electron
and the initial gluon, and $y_{\psi}$, $y_3$, $\phi_{\psi}$, and $\phi_3$
the rapidities and the azimuthal angles of the \J particle and the coproduced
gluon. Then, the fully differential cross section reads:
\begin{eqnarray}
&& d\sigma(ep\to e'\psi X)=
\frac{\alpha_s^2\,\alpha^2\,e_c^2\,|\R(0)|^2}{16\,\pi\,x_2\,s^2}\,
\frac{1}{4}\sum_{\mbox{{\tiny spins}}}\,
\frac{1}{8}\sum_{\mbox{{\tiny colours}}}
|{\cal M}(\g g\to\psi g)|^2  \nonumber \\ && \times
 \F_g(x_2,k_{2T}^2,\mu^2)\,\, \label{lips}
dk_{1T}^2\, dk_{2T}^2\, dp_{\psi T}^2\, dy_3\,dy_{\psi}\,
\frac{d\phi_1}{2\pi}\,\frac{d\phi_2}{2\pi}\, \frac{d\phi_{\psi}}{2\pi}.
\end{eqnarray}
The phase space physical boundary is determined by the inequality \cite{BycKaj}
\begin{equation}
G(\hat{s}, \hat{t}, k_3^2, k_1^2, k_2^2, m_{\psi}^2) \le 0,
\end{equation}
where $\hat{s}=(k_1+k_2)^2$, $\;\hat{t}=(k_1-p_{\psi})^2$, and $G$ is the
standard kinematic function \cite{BycKaj}.

The initial gluon momentum fractions $x_1$ and $x_2$ are calculated from
the energy-momentum conservation laws in the light cone projections:
\begin{eqnarray}
 (k_1+k_2)_{E+p_{||}}=
 x_1\sqrt{s} &=& m_{\psi T}\exp(y_{\psi})\;\; + \; |k_{3T}|\exp(y_3),
 \nonumber \\[-2mm] && \label{x1x2} \\[-2mm]
 (k_1+k_2)_{E-p_{||}}=
 x_2\sqrt{s} &=& m_{\psi T}\exp(-y_{\psi}) + |k_{3T}|\exp(-y_3), \nonumber
\end{eqnarray}
$m_{\psi T}=(m_{\psi}^2+|p_{\psi T}|^2)^{1/2}$.
Here, we preserve exact kinematics and do not neglect the ``small'' light-cone
component of the gluon momentum.
The multidimensional integration in Eq. (\ref{lips}) has been performed
by means of the Monte-Carlo technique, using the routine VEGAS \cite{VEGAS}.\\

\noindent {\bf  4~Numerical results and discussion}\\

We start the discussion by presenting a comparison between our theoretical
calculations and experimental data collected by the H1 collaboration at HERA
\cite{newH1}.
The collaboration reports on the measurement of a number of differential
cross sections: $d\sigma/dp_{T,\psi}^2$,  $d\sigma/dp_{T,\psi}^{*\,2}$,
$d\sigma/dz$, $d\sigma/dQ^2$, $d\sigma/dy$, $d\sigma/dy^*$, and $d\sigma/dW$,
where $p_{T,\psi}$ and $p_{T,\psi}^{*}$ are the \J transverse momenta
in the laboratory and $\gamma^*p$ center-of-mass systems, respectively,
$z$ is the \J inelasticity variable defined as $z=(p_\psi p_p)/(k_1 p_p)$,
$Q^2=-k_1^2$ is the photon virtuality, $y$ and $y^*$ are the \J rapidities
in the laboratory and $\gamma^*p$ systems, and $W$ is the $\gamma^*p$
invariant energy.
The data collected in the kinematic range 2 GeV$^2<Q^2<100$ GeV$^2$,
50 GeV $<W<$ 225 GeV, $0.3<z<0.9$, $p_{T,\psi}^{*\,2}>1$ GeV$^2$ will be
referred to as ``sample 1'' (see Fig. 2), while the data collected in the
kinematic range 12 GeV$^2<Q^2<100$ GeV$^2$, 50 GeV$<W<$225 GeV, $0.3<z<0.9$,
$p_{T,\psi}^{*\,2}>1$ GeV$^2$, $p_{T,\psi}^{2}>6.4$ GeV$^2$ will be referred
to as ``sample 2'' (see Fig. 3).

On the theoretical side, we have examined the
sensitivity of model predictions to the choice of the gluon distribution
functions, the renormalization scale in the strong coupling constant, the
value of the quark mass, and the values of the nonperturbative colour-octet
matrix elements.

The effect of the different equations (BFKL versus DGLAP) which govern the
evolution of gluon densities is found to be as large as a factor of 2 in the
production cross section. This is illustrated by a comparison of dash-dotted
and dashed histograms in Figs.~2 and 3, respectively. Note, however, that the
parametrizations which we have used here represent two extreme cases, so that
most of the other parametrizations available in the literature may be expected
to lie in between our curves (see \cite{Bo}, \cite{BJJPZ}).

A similar effect is connected with the renormalization scale $\mu^2$ in the
running coupling constant $\alpha_s(\mu^2)$. The calculations made with
$\mu^2=k_{2T}^2$ and $\mu^2=m_{\psi T}^2$ are represented by the dash-dotted
and dotted histograms in Figs.~2 and 3. Note that the setting $\mu^2=k_{2T}^2$
is only possible in the \k approach, while it is meaningless in the collinear
calculations where the parton transverse momentum is neglected:
$k_{2T}\equiv 0$.

The quark mass plays in the calculations two essentially different roles.
The ``current'' mass $m_c$ is present in the expressions for the perturbative
matrix elements (\ref{Ma})-(\ref{Mb}). The ``constituent'' mass determines the
phase space of the reaction via its connection to the physical mass of the
final state, $m_{\psi}=2m_c$. However, it is worth pointing out that this
connection is not strict. In fact, it may be violated by the effects of binding
energy and internal quark motion \cite{Wyler}. The sensitivity of model
predictions to the quark mass setting was examined in a toy calculation, where
the ``current'' mass $m_c$ and the ``constituent'' mass $m_{\psi}/2$ were
treated as independent parameters. We found that the production cross section
is only sensitive to the ``constituent'' mass but remains rather stable against
variations in the ``current'' mass. In the rest of the paper we will be always
using $m_c=m_{\psi}/2=1.55$ GeV.

The contributions from the $2\to 1$ colour-octet subprocesses are cut away
by the experimental restriction $z<0.9$.
Turning to the $2\to 2$ colour-octet contributions, one has to take care
about the infrared instability of the relevant matrix elements. In a rigorous
approach, one has to consider the corresponding $2\to 1$ subprocesses at
next-to-leading order. Then, the interference between the LO and NLO
contributions must cancel the divergent parts of the $2\to 2$ subprocesses.
Such calculations have been performed in the collinear factorization in
\cite{Petr}. Since the corresponding results are not yet available in the
$k_T$-facctorization, we use an approximate phenomenological approach. In
order to restrict the $2\to 2$ subprocesses to the perturbative domain, we
introduce the regularization parameter $q^2_{\mbox{reg}}$, so that all
propagators are kept away from their poles by a distance not less than
$q^2_{\mbox{reg}}$. It may be argued
that the nonperturbative parts of the $2\to 2$ subprocesses can be absorbed
into $2\to 1$ subprocesses; that is, when the emitted gluon is soft,
one can consider the final state as represented by a single particle rather
than by two. This our suggestion is similar to the one made in papers
\cite{BraFle}-\cite{Braat}, \cite{gfrag} in the usual factorization scheme.
In this approach, the regularization parameter $q^2_{\mbox{reg}}$
in the $2\to 2$ processes and the nonperturbative colour-octet matrix elements
in the $2\to 1$ processes must be correlated \cite{gfrag} to avoid double
counting between the hard and soft gluons in the final state (and so, to
avoid sensitivity of the results to the choice of $q^2_{\mbox{reg}}$).
The numerical results shown in Figs. 2 and 3 are obtained with setting
$q^2_{\mbox{reg}}=1$~GeV$^2$ (which may be regarded as the lower limit
of the perturbative domain).

The results shown in Figs.~2 and 3 are obtained with the nonperturbative
colour-octet matrix elements of Ref. \cite{ChoLei}. If the values exracted
from the recent analysis \cite{j_sha} were used unstead, the contribution
from the colour-octet states would be a factor of 5 lower\footnote{%
  The nonperturbative matrix elements given in Ref. \cite{ChoLei} are as
  follows: $|{\cal R}_{^3S_1^1}(0)|^2=8\times 10^{-1}$ GeV$^3$,
           $|{\cal R}_{^1S_0^8}(0)|^2=8\times 10^{-3}$ GeV$^3$,
           $|{\cal R}_{^3S_1^8}(0)|^2=8\times 10^{-3}$ GeV$^3$,
          $|{\cal R}'_{^3P_0^8}(0)|^2=7\times 10^{-3}$ GeV$^5$.
  According to \cite{j_sha}, the $\langle^1\!S_0^8\rangle$ and
  $\langle^3\!P_J^8\rangle$ matrix elements should be reduced by a factor
  of 5, while the $\langle^3\!S_1^8\rangle$ matrix element should be reduced
  by at least a factor of 50 or enven set to 0. This violates the naive
  nonrelativistic QCD scaling rules, but is consistent with estimates obtained
  within the \k approach by other athors \cite{Teryaev}, \cite{Chao2}.}.
Irrespective to the particular choice of the nonperturbative matrix elements,
the production of \J particles at the conditions under study is not dominated
by the colour-octet mechanism.
This conclusion is also supported by an independent analysis \cite{Artem}
of recent ZEUS photoproduction data \cite{ZEUS1}.
In general, the data are consistent with the predictions of the colour-singlet
model and lie within the theoretical band provided by reasonable variations in
model parameters. Although no need is seen in the colour-octet contributions,
their presence does not lead to serious discrepancies in the visible kinematic
area. 

We find it instructive to proceed the discussion by presenting a comparison
with another data displayed by the H1 collaboration in their earlier
publication \cite{oldH1}.
The data collected in the range 2 GeV$^2<Q^2<80$ GeV$^2$, 40 GeV $<W<$ 180 GeV,
$z>0.2$ constitute two correlated samples.
In the ``inelastic'' sample, an additional cut on the final state hadron mass
was applied, $M_X>10$ GeV, while the sample without any additional cuts is
referred to as ``inclusive''.

Although these data are, probably, less convincing in statistics, they are
more useful and interesting from the theoretical point of view.
Firstly, in the inclusive production sample, the high-$z$ kinematic range
$z\simeq 1$ provides an access to the $2\to 1$ colour-octet contributions.
Secondly, the cut on the final state hadron mass $M_X$ in the inelastic
sample prevents the $2\to 2$ colour-octet matrix elements from infrared
divergence.

In Fig.~4, we present a comparison between the data and our predictions for
the inelastic production. The final state hadron mass $M_X$ is calculated as
the invariant mass of the proton remnant and the final state gluon produced
in the $2\to 2$ subprocesses (\ref{psig})-(\ref{3PJg}). The requirement that
the final state mass $M_X$ be large means that the final state gluon must be
hard. Thus, the theoretical estimates of the colour octet contributions are no
longer dependent on the artificial regularization parameter $q^2_{\mbox{reg}}$.
One can see that the data are reasonably described within the colour-singlet
production mechanism, and the colour-octet contributions are not needed.
Moreover, the colour-octet contributions to the high-$z$ region look even
superfluous. The overall situation seems to favour the low values of the
nonperturbative matrix elements proposed in \cite{j_sha}.

In the inclusive production sample (see Fig.~5), a significant deficit is seen
in the $z$-distribution at $z\simeq 1$, although the intermediate values of $z$
are described quite well. In the \J transverse momentum spectrum, the deficit
is only seen at low $p_{T,\psi}$, while there is no discrepancy at higher
$p_{T,\psi}$ values. These properties may be taken as an indication that the
inclusive sample contains large contributions from diffractive processes.
As a consequence, our calculations underestimate the absolute \J production
rate by a factor of 3.

We conclude the discussion by presenting our results on the \J spin alignment
in Fig.~6. The data are expressed in terms of the leptonic decay angular
parameter $\alpha$, which characterizes the azimuthal angle distribution
measured in the \J rest frame with respect to a given reference axis:
$d\Gamma_{ll}\sim 1+\alpha\cos{(\theta)}$. The cases $\alpha=1$ and
$\alpha=-1$ correspond to transverse and longitudinal polarizations
of the \J meson, respectively. Our calculations show that the fraction
of longitudinally polarized mesons increases with increasing $Q^2$, which
is a consequence of the enhancement of the longitudinal component in the
polarization vector of a virtual photon.
Unfortunately, the experimental data are rather indefinitive.
In this case, we can derive no conclusions
on the agreement or disagreement between the theory and experiment.

In the situation which we consider, the effects of the gluon off-shellness are
shadowed by the effects of highly virtual photons and are not clearly visible.
At a different situation, when either the initial photons are real or the
process is mediated by gluons only, the \J polarization properties are found
to be perfectly compatible with the predictions of the \k approach, as it was
demonstrated earlier in Refs. \cite{jbfkl}, \cite{Artem} and \cite{j_sha}
devoted to the analysis of \J photoproduction and hadroproduction data.\\

\noindent {\bf  5~Conclusion}\\

Here we have addressed the issue of performing a global analysis of quarkonium
electroproduction within the $k_T$-factorization approach. The state of the art
has not yet reached the precise quantitative level. There are uncertainties
connected with the choice of unintegrated gluon densities, the renormalization
scale in the strong coupling constant, the inclusion of next-to-leading-order
subprocesses, and the nonperturbative colour-octet transitions.

At the same time, the \k approach shows a number of important achievements.
As a general feature, the model behavior is found to be perfectly compatible
with the available data on the production of various quarkonium states at
modern colliders. The model succeeds in describing the $p_T$-spectra of
$J/\psi$, $\chi_c$, and $\Upsilon$ mesons at the Fermilab Tevatron and provides
a consistent picture of the production of \J mesons by real and virtual photons
at HERA (as it was demonstrated earlier in Refs. \cite{jbfkl}, \cite{Artem} and
\cite{j_sha}).
The model even succeeds in describing the polarization phenomena observed
in both $pp$ and $ep$ interactions, thus providing an important insight for
solving a long-term puzzle.

As a result of the complex interplay of all theoretical uncertainties, the
numerical analysis becomes rather umbiguous. The electroproduction data can
be successfully described within the leading-order colour-singlet mechanism
alone. No need is seen in the colour-octet contributions. At the same time,
there are no serious contradictions with the data if the colour-octet
contributions are included. The effects of the different unintegrated
gluon distributions can hardly be separated from the effects connected
with the running coupling constant.\\

\newpage
\noindent{\bf Acknowledgments}\\

One of us (N.Z.) was supported in part by Russian Foundation for Basic
Research under grant RFBR  02-02-17513.

\newpage
\begin{figure} \label{fig1} \caption{%
 Feynman graphs representing the photon-gluon fusion mechanism in the
 colour-singlet and colour-octet models. Only the perturbative skeleton
 is presented; the soft gluons corresponding to the nonperturbative
 colour-octet transitions are not shown.}
\end{figure}
\begin{figure} \label{samp1}\caption{%
A comparison between the theoretical predictions and experimental data
\cite{newH1} in the kinematic range $2<Q^2<100$ GeV$^2$, $50<W<225$ GeV,
$0.3<z<0.9$, $p_{T,\psi}^{*\,2}>1$ GeV$^2$.~~
Dash-dotted histogram, the colour-singlet contribution
with BFKL gluon density and $\alpha_s(k_{2T}^2)$;~~
dashed histogram, the colour-singlet contribution
with DGLAP gluon density and $\alpha_s(k_{2T}^2)$;~~
dotted histogram, the colour-singlet contribution
with BFKL gluon density and $\alpha_s(m_{T,\psi}^2)$;~~
solid histogram, the sum of the colour-singlet and colour-octet contributions, 
with BFKL gluon density, $\alpha_s(k_{2,T}^2)$, colour-octet matrix elements 
as in \cite{ChoLei}, and $q^2_{\mbox{reg}}=1$~GeV$^2$.}
\end{figure}
\begin{figure} \label{samp2}\caption{%
A comparison between the theoretical predictions and experimental data
\cite{newH1} in the kinematic range $12<Q^2<100$ GeV$^2$, $50<W<225$ GeV,
$0.3<z<0.9$, $p_{T,\psi}^{*\,2}>1$ GeV$^2$, $p_{T,\psi}^{2}>6.4$ GeV$^2$.~~
Dash-dotted histogram, the colour-singlet contribution
with BFKL gluon density and $\alpha_s(k_{2T}^2)$;~~
dashed histogram, the colour-singlet contribution
with DGLAP gluon density and $\alpha_s(k_{2T}^2)$;~~
dotted histogram, the colour-singlet contribution
with BFKL gluon density and $\alpha_s(m_{T,\psi}^2)$;~~
solid histogram, the sum of the colour-singlet and colour-octet
contributions, with BFKL gluon density, $\alpha_s(k_{2,T}^2)$,
colour-octet matrix elements as in \cite{ChoLei},
and $q^2_{\mbox{reg}}=1$~GeV$^2$.}
\end{figure}
\begin{figure} \label{inel}\caption{%
A comparison between the theoretical predictions and experimental data
\cite{oldH1} on the inelastic \J production in the range $2<Q^2<80$ GeV$^2$,
$40<W<180$ GeV, $z>0.2$, $M_X>10$ GeV.~~
Dash-dotted histogram, the colour-singlet contribution
with BFKL gluon density and $\alpha_s(k_{2T}^2)$;~~
dashed histogram, the colour-singlet contribution
with DGLAP gluon density and $\alpha_s(k_{2T}^2)$;~~
dotted histogram, the colour-octet contribution
with BFKL gluon density, $\alpha_s(k_{2T}^2)$ and
colour-octet matrix elements as in \cite{ChoLei};~~
solid histogram, the sum of the colour-singlet and colour-octet contributions.}
\end{figure}
\begin{figure} \label{incl}\caption{%
A comparison between the theoretical predictions and experimental data
\cite{oldH1} on the inclusive \J production in the range $2<Q^2<80$ GeV$^2$,
$40<W<180$ GeV, $z>0.2$.~~
Dash-dotted histogram, the colour-singlet contribution
with BFKL gluon density and $\alpha_s(k_{2T}^2)$;~~
dashed histogram, the colour-singlet contribution
with DGLAP gluon density and $\alpha_s(k_{2T}^2)$;~~
dotted histogram, the colour-octet contribution
with BFKL gluon density, $\alpha_s(k_{2T}^2)$,
colour-octet matrix elements as in \cite{ChoLei},
and $q^2_{\mbox{reg}}=1$~GeV$^2$;~~
solid histogram, the sum of the colour-singlet and colour-octet contributions.}
\end{figure}

\begin{figure} \label{angle}\caption{%
A comparison between the theoretical predictions and experimental data
on the \J spin alignment represented in terms of the decay lepton angular
distributions.
Left panel,
      kinematic range 1 (as in Fig.~2) \cite{newH1}, $Q^2<6.5$ GeV$^2$;~~
right panel,
      kinematic range 1 (as in Fig.~2) \cite{newH1}, $Q^2>6.5$ GeV$^2$.}
\end{figure}
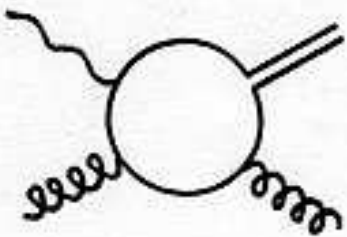
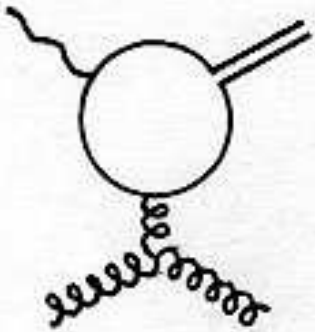
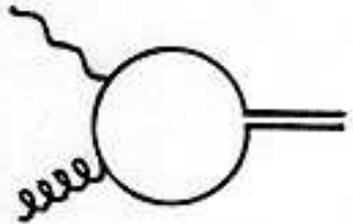
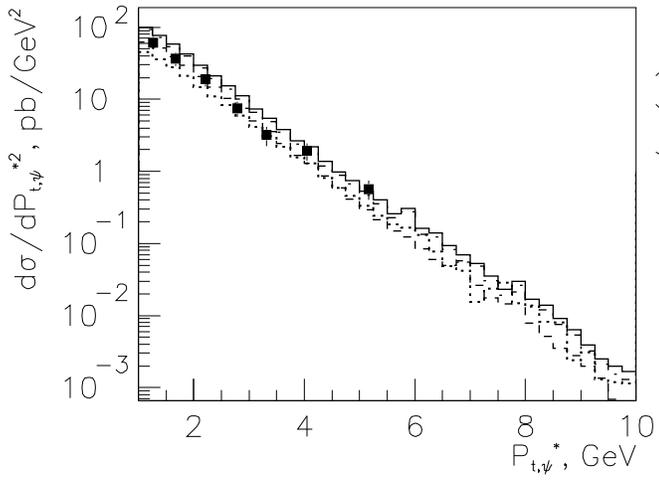
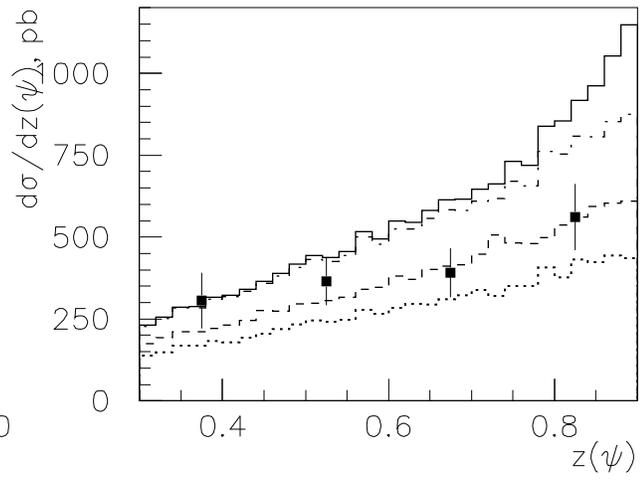
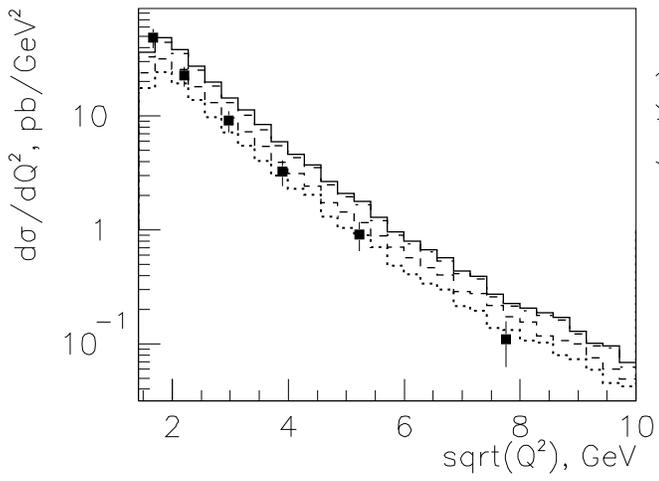
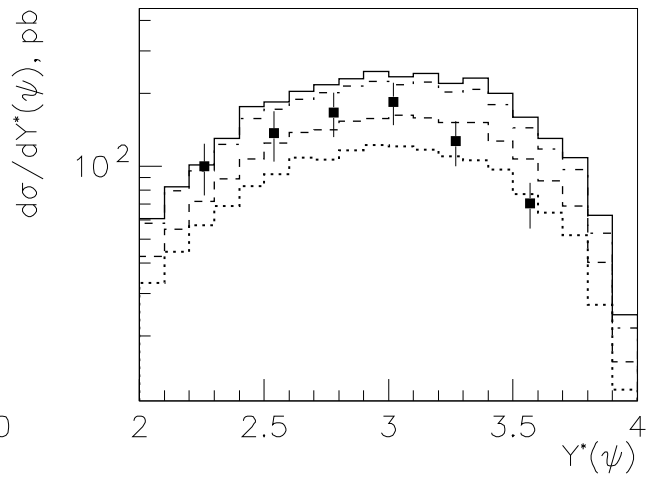
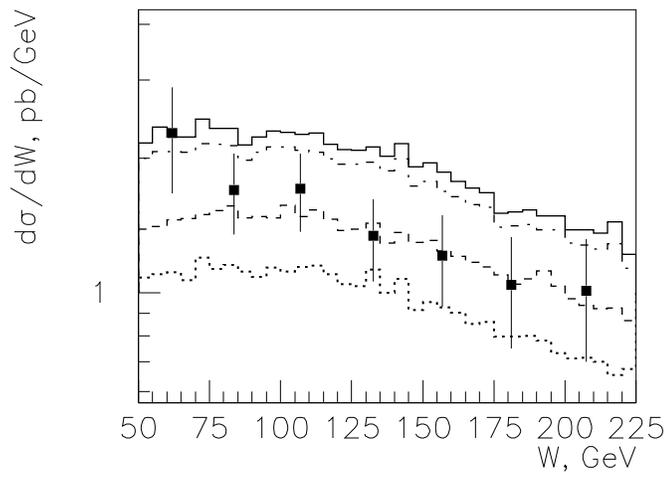
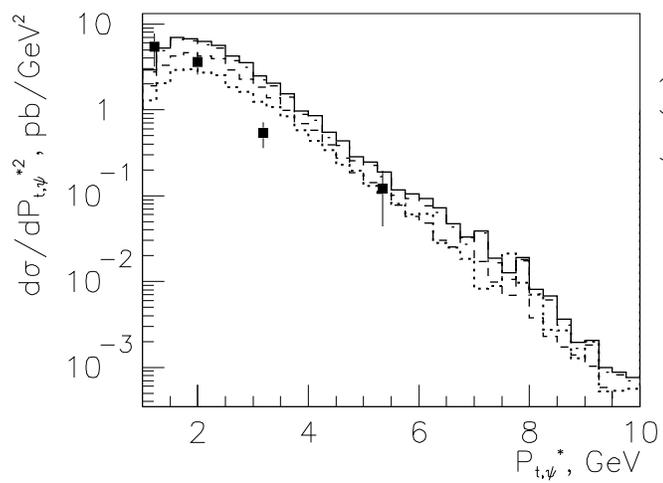
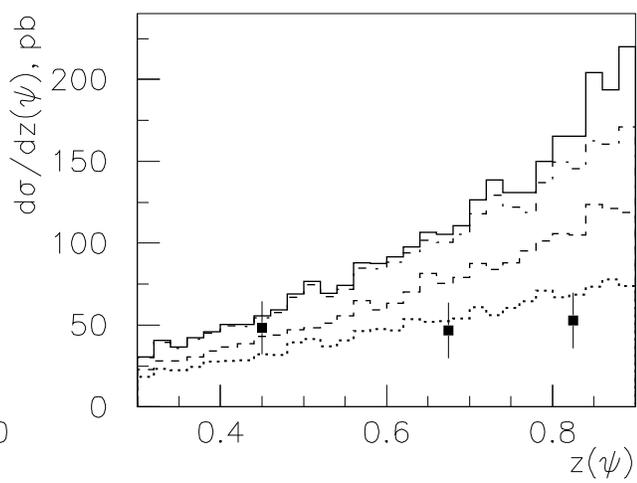
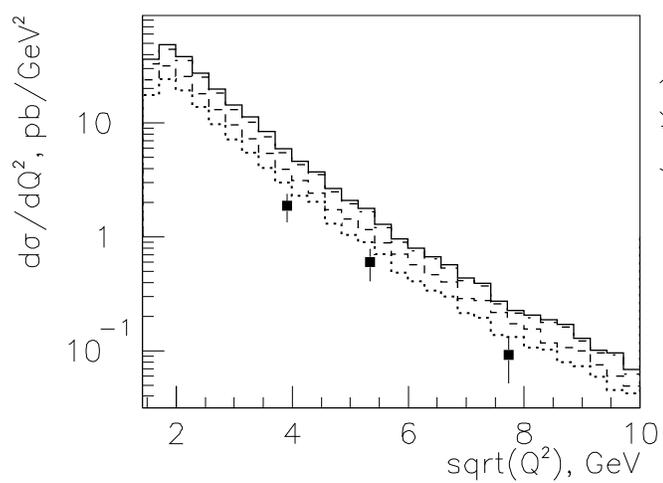
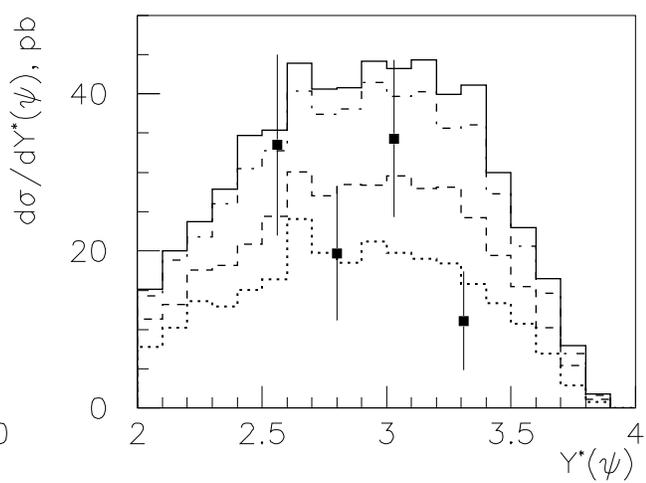
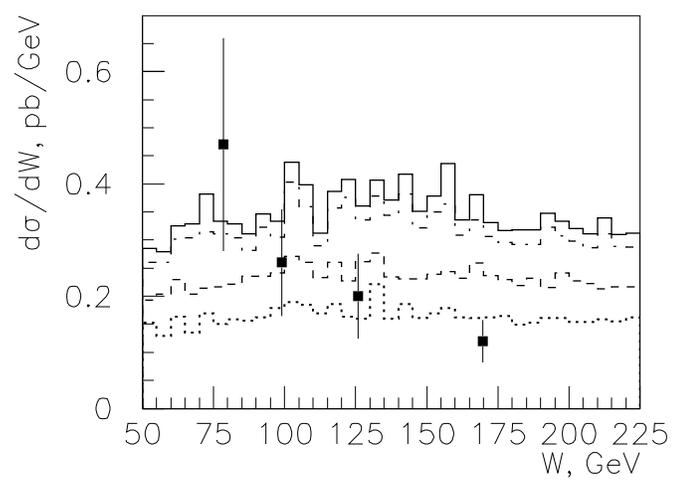
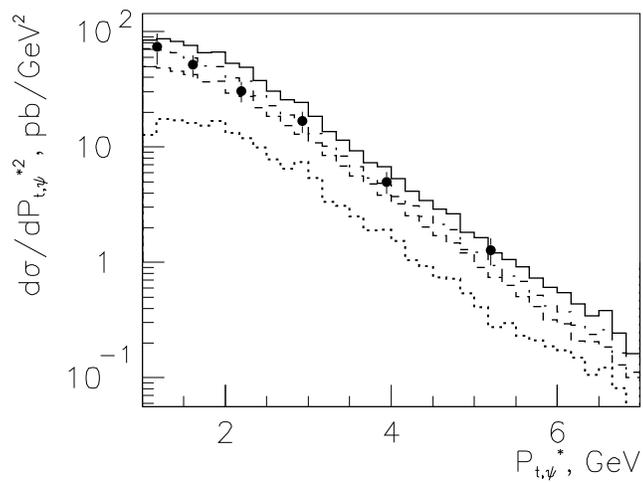
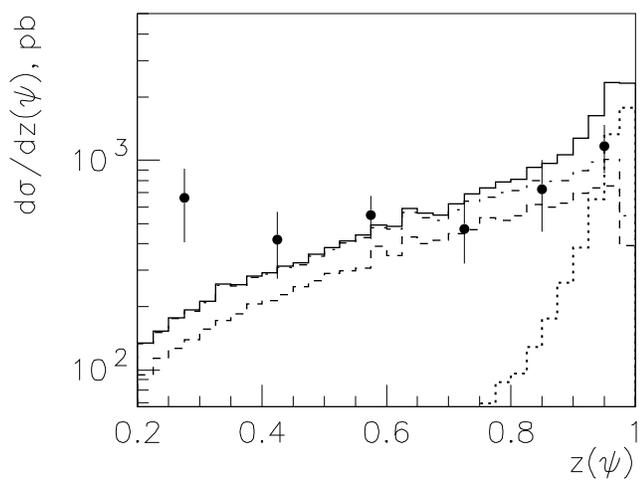
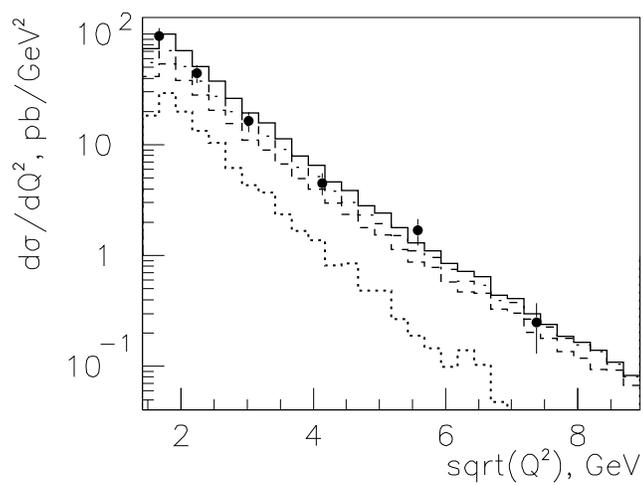
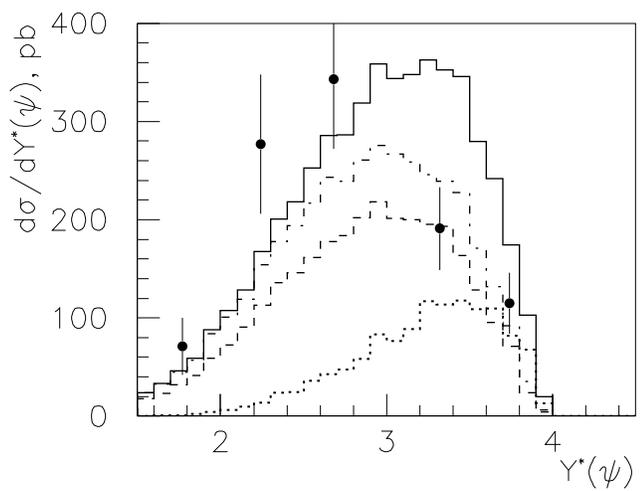
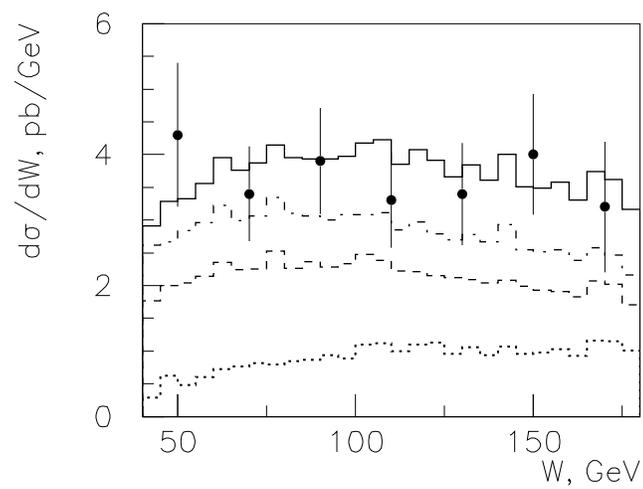
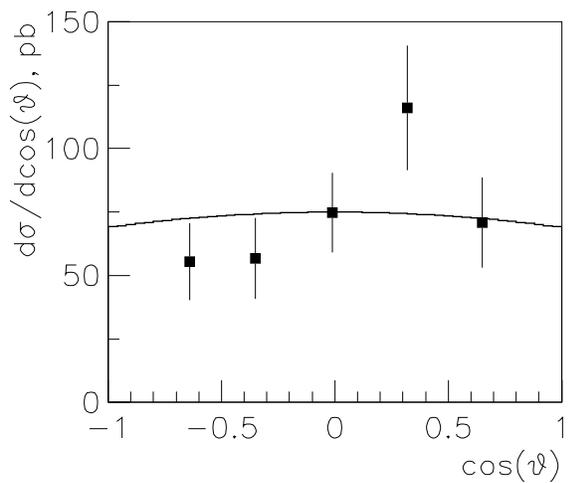
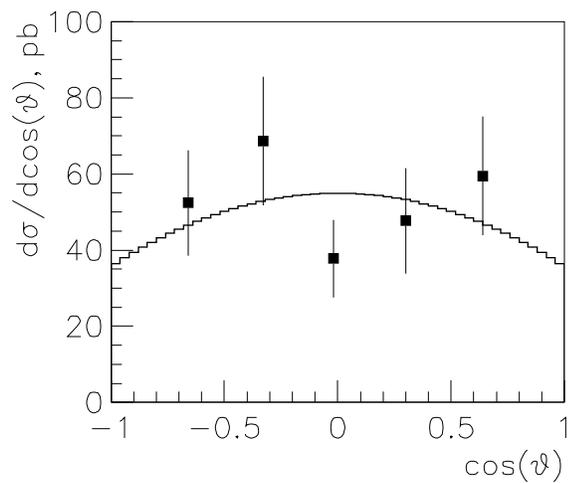
\end{document}